\documentclass[prl, twocolumn]{revtex4}
\usepackage{dcolumn}
\usepackage{graphicx}
\begin{document}

\title{Non-Markovian fluctuations in Markovian models of protein dynamics}
\author{Arti Dua$^1$ and R. Adhikari$^2$}
\affiliation{$^{1}$Department of Chemistry, Indian Institute of Technology, Madras, Chennai 600036, India\\
				$^{2}$The Institute of Mathematical Sciences, CIT Campus, Chennai 600113, India}

\date{\today}

\begin{abstract}
Recent experiments using fluorescence spectroscopy have been able to probe the dynamics of conformational fluctuations in proteins. The fluctuations are Gaussian but do not decay exponentially, and are therefore, non-Markovian. We present a theory where non-Markovian fluctuation dynamics emerges naturally from the superposition of the Markovian fluctuations of the normal modes of the protein. A Rouse-like dynamics of the normal modes provides very good agreement to the experimentally measured correlation functions. We provide simple scaling arguments rationalising our results. 
\end{abstract}


\maketitle

Proteins molecules are the buiding blocks of life. The three-dimensional physical structure of a protein is intimately related to its biological function. Proteins function in an environment where noise is ubiquitous. The three-dimensional conformation of a protein, therefore, is not static but itself undergoes fluctuations. Since protein function is so strongly determined by protein structure, the fluctuating nature of a protein molecule has important implications for its biological function \cite{lu:1998, oijen:2003}. A precise determination of the static and dynamic properties of  conformational fluctuations in proteins is, therefore, of great importance. 

In recent experiment \cite{yang:2003}, conformational fluctuations of the protein flavin reductase have been observed and characterised using single-molecule fluorescence spectroscopy. The fluorescence lifetime is directly correlated to the distance between the flavin (fluorophore) and tyrosine (quencher) groups within the protein. The distance fluctuations between the flavin and tyrosine groups, then, gives an indirect measure of the fluctuations of the entire protein. Remarkably, the experiments find that the fluctuations remain correlated over a five decades in time, spanning the range of $10^{-4}s$ - $1s$. This is indicative of the presence of multiple relaxation mechanisms operating at different time scales. Further, there is convincing evidence that the fluctuations are Gaussian, and when taken together with the absence of single-exponential decay of the correlations, imply that they are also non-Markovian. 

In this Letter, we show how  a Markovian dynamics for the  protein normal modes generically produces non-Markovian fluctuations in the distance between two residues on the protein backbone. Imposing the simplest Rouse-like dynamics for the normal modes we obtain all correlation functions for the distance fluctuations and find them to be in very good agreement with experiments \cite{yang:2003, kou:2004, min:2005}. Simple scaling arguments are provided to rationalise our analytical calculations. 

Before presenting our detailed calculation, we illustrate the basic mechanism by which non-Markovian behaviour arises in this problem. Consider the Ornstein-Uhlenbeck process (OUP) which describes the velocity $v(t)$ of Brownian motion. This is a stationary, Gaussian, Markovian process with a correlation function $\rho_0(\tau) = \langle v(t)v(t+\tau)\rangle = k_BT\exp(-\Gamma \tau)$, where $\Gamma^{-1}$ is the relaxation time. Take two such uncorrelated processes $v_1(t)$ and $v_2(t)$, each with distinct relaxation times $\Gamma_1$ and  $\Gamma_2$,  and ask for properties of the stochastic process described by their sum $u(t) = v_1(t) + v_2(t)$. Since each $v_i$ is Gaussian and stationary, so is their sum $u$. The correlation function of the sum is $\rho(\tau) = \langle u(t)u(t+\tau)\rangle = \langle (v_1(t) + v_2(t))(v_1(t+\tau)+v_2(t+\tau))\rangle$,  and since the processes are uncorrelated,  is the sum of the correlation functions of the individual processes, $\rho(\tau) = k_BT[\exp(-\Gamma_1\tau) + \exp(-\Gamma_2\tau)]$. Then, from Doob's theorem \cite{vankampen:1981}, which says that a Gaussian, stationary process is Markovian if and only if its correlation function is a single exponential, we conclude that $u(t)$ is Gaussian, stationary, but $\emph{non-Markovian}$. Generalising, an abitrary superposition of $N$ distinct but uncorrelated OUPs, $u(t) = \sum_{i=1}^N\alpha_i v_i(t)$, is also Gaussian, stationary, and non-Markovian. Thus, non-Markovian behaviour can arise very generally from a superposition of Markovian processes. It precisely this mechanism which, as we show below, generates the non-Markovian fluctuations seen in protein dynamics.

A minimal model of a protein replaces the complicated stereochemistry of the amino acids and its associated secondary and tertiary structures by a simple connected chain of beads and springs \cite{banavar:2005}. The relaxations in such a model derive from interactions between the beads and the combined effect of fluctuations and dissipation due to the solvent. In the energetic ground state, the conformation is labelled by the positions ${\bf R}^0_n$  of the beads, where the subscript $n$ is a position label along the chain. Independent of the specific nature of the interactions, conformational fluctuations about the ground state can be described by a parametrisation ${\bf R}_n = {\bf R}^0_n + {\bf u}_n$. Here, ${\bf u}_n$ is the deviation of the $n-$th bead from its ground state conformation. Then, the instantaneous distance ${\bf d}_{mn}(t)$ between two monomers located at $m$ and $n$ is given by ${\bf d}_{mn}(t) = {\bf R}_m(t) - {\bf R}_n(t) = {\bf d}_{mn}^0 + {\bf u}_m(t) - {\bf u}_n(t)$, where ${\bf d}_{mn}^0 = {\bf R}_m^0 - {\bf R}_n^0$ is the equilibrium distance. For a chain whose ends are not tethered and therefore free of external forces, the displacements must satisfy $\partial {\bf u}_n/ \partial n = 0$ at $n=0$ and $n=N$. This motivates the introduction of normal modes of the form ${\bf u}_{n}(t) = 2 \sum_{p=1}^{\infty} {\bf Q}_p(t)\cos(p\pi n/N)$, in terms of which the distance is
\begin{equation}
{\bf d}_{mn}(t) = {\bf d}_{mn}^0 + 2 \sum_{p=1}^{\infty} {\bf Q}_p(t) [ \cos(p\pi m/N) - \cos(p \pi n/N)].
\end{equation}
This key equation shows that the distance fluctuations are linearly related to the fluctuations of the normal modes. If the normal mode fluctuations are Gaussian and Markovian, the distance fluctuations, by our previous argument, are generically non-Markovian. 

The simplest model of polymer dynamics which yields a Gaussian and Markovian fluctuation for the normal modes is the Rouse model \cite{doi:1986} . Here, we impose Rouse-like dynamics on the harmonic deviations ${\bf u}_n$, 
\begin{equation}
\zeta\frac{\partial{\bf u}_n(t)}{\partial t} = \frac{3 k_B T}{b^2} \frac{\partial^2{\bf u}_n(t)}{\partial n^2} + {\bf f}_n(t), 
\end{equation}
so that the overdamped dynamics of the fluctuations is a balance between the frictional force proportional to $\zeta$ times the velocity of the $n-$th monomer, an entropic restoring force proportional to $3 k_B T/b^2$ which arises due to the connectivity of the chain, and the injection of thermal fluctuations from the solvent. The constant term in the Rouse mode expansion describing the motion of the center of mass of the chain and has been ignored here since it does not enter the expression for the distance.  The dynamics of the Rouse modes follows immediately as
\begin{equation}
\zeta_p\frac{\partial{\bf Q}_{p}(t)}{\partial t} = - k_p{\bf Q}_p(t)+  {\bf F}_{p}(t), 
\end{equation}
where $\zeta_p = 2 N \zeta$, $k_p = 6 p^2 \pi^2 k_B T/N b^2$ and ${\bf f}_{n}(t) = 2 \sum_{p=1}^{\infty} {\bf F}_p(t)\cos(p\pi n/N)$. 
The fluctuations of the Rouse modes are, therefore, identical to the fluctuations of the velocity of a Brownian particle, both being governed by the OUP. The correlations between the modes is given by \cite{doi:1986}
\begin{equation}
\left<{\bf Q}_p(t)\cdot {\bf Q}_q(t + \tau)\right> = \delta_{pq}\frac{N b^2}{(p^2+q^2)\pi^2} \exp(-p^2 \tau/\tau_1), 
\end{equation}
showing that each Rouse mode has a distinct relaxation time and is unocorrelated with every other Rouse mode. Here, $\tau_1 = Nb^2\zeta_p/6\pi^2k_BT$ is the relaxation time of the first Rouse mode. 

Combining the results of Eq. 1, and Eq. 4, we see that ${\bf d}_{mn}(t)$ is a stochastic process which is an infinite superposition of OUPs. Explicitly, the correlation function $\rho_{mn}(\tau) = \langle {\bf d}_{mn}(t)\cdot {\bf d}_{mn}(t+\tau)\rangle$ of this process is 
\begin{equation}\label{eq:corr}
\rho_{mn}(\tau) = 2 \sum_{p=1}^{\infty}\frac{N b^2}{p^2\pi^2}  [ \cos(p\pi n/N) - \cos(p \pi m/N)]^2 e^{-p^2 \tau/\tau_1}.
\end{equation}
By Doob's theorem, it is immediately clear that the dynamics of ${\bf d}_{mn}(t)$ is non-Markovian. Since ${\bf d}_{mn}(t)$ is a Gaussian process, all higher order time correlation functions can be expressed in terms of the $\rho_{mn}(\tau)$ using Wick's theorem.  We call the stochastic process defined by Eq. 1 and Eq. 3 the superposed Ornstein-Uhlenbeck process. The two-point correlation $\rho_{mn}(\tau)$ completely specifies the process. 

Our work thus far is, in a formal sense,  a Markovian embedding (in terms of the normal modes) of a non-Markovian process (the distance fluctuations). Such Markovian embeddings are also used in describing the underdamped dynamics of a Brownian particle in a potential. The stochastic process describing the position alone is non-Markovian, but the joint process in the enlarged set of  position and velocity variables is Markovian \cite{vankampen:1981}. In the present case, the Markovian embedding is also Gaussian, and it is this simplification that allows us to calculate all correlation functions in terms of the two-point correlation  $\rho_{mn}(\tau)$.

We now turn to comparing our analytical results with data from the experiments \cite{yang:2003, kou:2004, min:2005}. The experiments measure the fluorescence liftetime $\gamma^{-1}(t)$ which is related to the distance $d_{mn}(t) =\sqrt{{\bf d}_{mn}(t)\cdot{\bf d}_{mn}(t)}$ between the fluorophore and 
the quencher as
\begin{equation}
\gamma(t) = k_0e^{-\lambda d_{mn}(t)}
\end{equation}
where $k_0$, $\lambda$ are parameters determined by the protein, ${\bf d}_{mn}(t) = {\bf d}_{mn}^0 + {\bf u}_m(t) - {\bf u}_n(t)$ and $d_{mn}^0$ is the mean value of the distance. Then correlation functions of the lifetimes $\delta\gamma^{-1}(t) = \gamma^{-1}(t) -\langle\gamma^{-1}\rangle$  are related to correlations in the distance fluctuations. This relation is simple for the two-point correlation function \cite{kou:2004}, 
\begin{equation}
\langle \delta\gamma^{-1}(t)\delta\gamma^{-1}(t + \tau)\rangle = k_0^{-2} e^{2\beta d^0_{mn} + \beta^2\rho_{mn}(t)}(e^{\beta^2\rho_{mn}(\tau)} -1 )
\end{equation}
but becomes complicated for three- and higher-point correlations. Explicit forms for three and four point correlations are given in \cite{kou:2004}.

We compare the results for two- and four-point fluorescence lifetime correlations using the correlation functions of the superposed OUP. The experimentally known values of the parameters are $d_{mn}^0 = 4.5\AA$, $\beta = 1.4 \AA^{-1}$, and $\gamma/k_BT = 0.48 \AA^{-2}s$ \cite{kou:2004}. With fitting parameters which are very close to these estimates,
the agreement between the theoretical prediction and the experimental data for the two-point correlation is good over the entire $5$ decades in time,  as shown in Fig. 1. In Fig. 2 we compare theory and experiment for the four-point function by fitting the same parameters used in Fig. 1.
Again, the agreement is good over the full $5$ decades in time. 
\begin{figure}
\includegraphics[width=8cm]{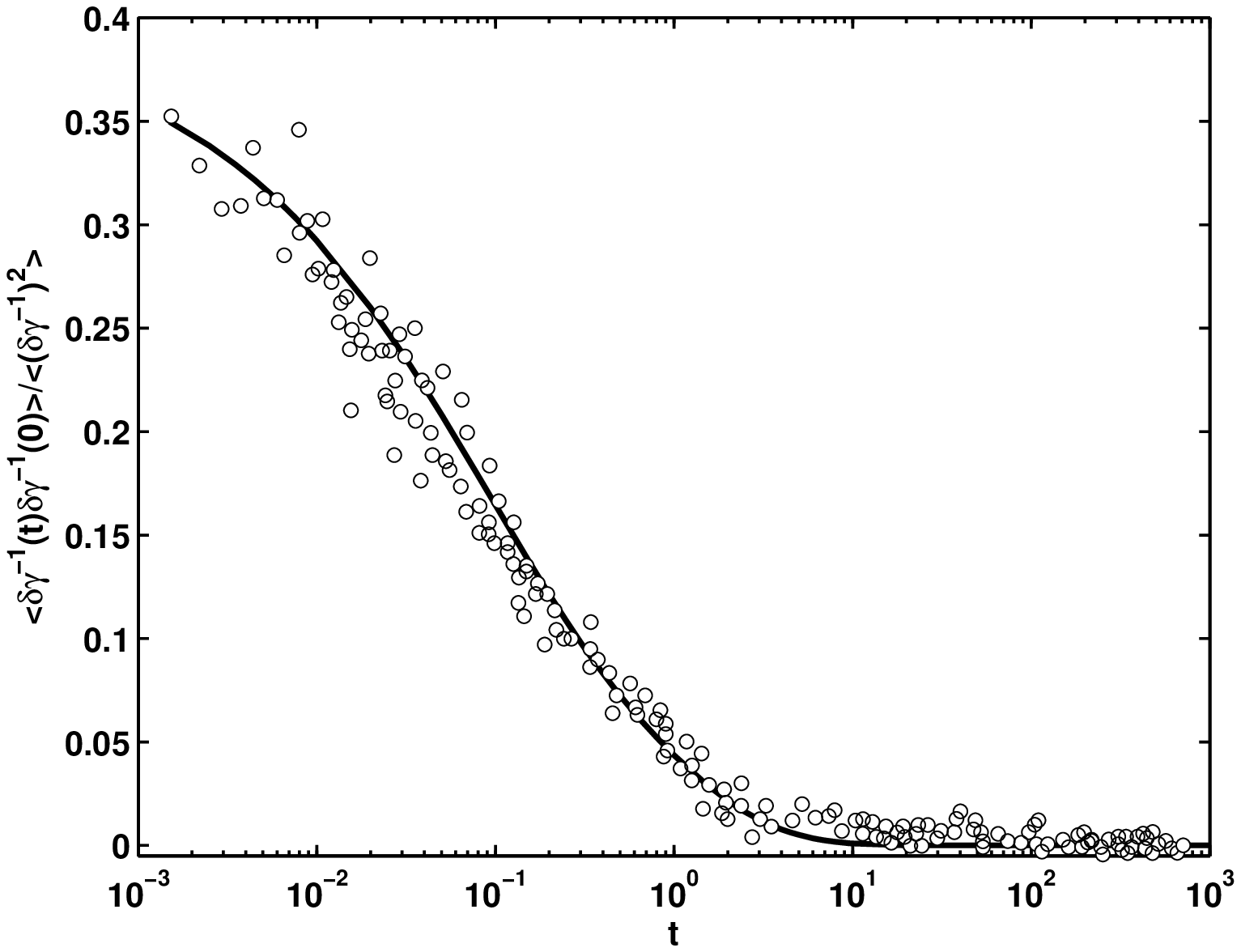}
\includegraphics[width=8cm]{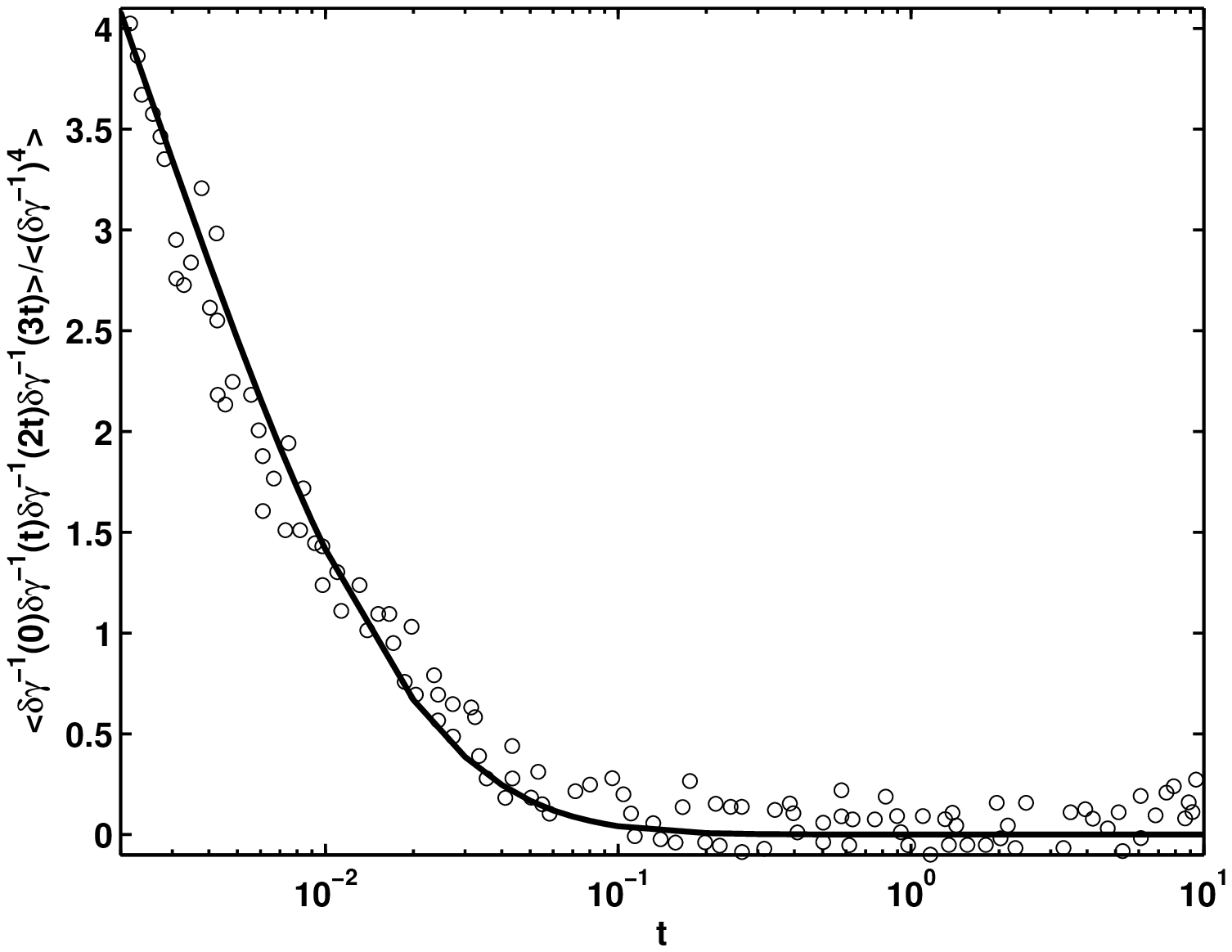}
\caption{\label{fig:twpoint} The normalised two-point (above) and four-point (below) autocorrelation functions of the fluorescence liftetimes plotted against time $t$ in seconds. The solid line is the theoretical curve obtained by using the the normalized correlation function in Eq.~\ref{eq:corr} while the data is from Ref.~\cite{kou:2004}. The theoretical curve has been multiplied by the first data point. The parameters used are $\beta = 1.3 \AA^{-1}$, $d^0_{mn} = 2.8\AA$, $\gamma/k_BT = 0.4\AA^{-2}s$, $m=6$, $n=30$ and $N=30$.}
\end{figure}

The interesting symmetry in time of the three-point correlation function $\langle \delta\gamma^{-1}(0)\delta\gamma^{-1}(t_1)\delta\gamma^{-1}(t_1 + t_2)\rangle = \langle\delta\gamma^{-1}(0)\delta\gamma^{-1}(t_2)\delta\gamma^{-1}(t_1 + t_2)\rangle$ observed in experiment \cite{kou:2004} holds {\it by construction} for the superposed OUP. This is so because, by stationarity, all times indices in the first term of the identity can be shifted by $t_1 + t_2$, which reproduces the second term of the identity. Similar relations also hold for the higher-point correlation functions. 

The Rouse-like dynamics of the normal modes, then, can capture all the essential features of the experimental fluctuation spectrum both qualitatively and quantitatively. This provides evidence that the large-scale long-time behaviour of proteins are identical to those of structureless Rousian polymer chains, not only for statics \cite{banavar:2005}, but also for aspects of dynamics.

To fully specify the superposed OUP we must obtain all the joint probabilities  $P({\bf d}, t; {\bf d^{\prime}}, t^{\prime}; {\bf d^{\prime\prime}}, t^{\prime\prime};\ldots)$ on the sample paths ${\bf d}_{mn}(t) = {\bf d}$ \cite{vankampen:1981}. As the process is Gaussian, all such probabilities are multivariate Gaussian distributions determined by the single function $\rho_{mn}(\tau)$\cite{fox:1978}. The one-point distribution $P({\bf d}, t)$ is independent of time by stationarity and is a Gaussian in ${\bf d}$ with mean ${\bf d}_{mn}^0$ and variance $\rho_{mn}(0) = Nb^2$. The two-point distribution is 
\begin{equation}
P({\bf d}, t; {\bf d}^{\prime}, t^{\prime}) = {1\over (2\pi)^3 \det[{\bf C({\tau})}]^{{1\over 2}}}\exp[-{1\over 2}\Delta_{i}C^{-1}_{ij}({\tau})\Delta_{j}]
\end{equation}
where  $t^{\prime} = t + \tau$,  ${\bf \Delta} = (d_x, d_y, d_z, d_x^{\prime}, d_y^{\prime},d_z^{\prime})$, and $C_{ij}(\tau) = \langle \Delta_i(t)\Delta_j(t + \tau)\rangle = \int_{\Delta_i, \Delta_j} \Delta_i \Delta_j P({\bf d}, t; {\bf d}^{\prime}, t^{\prime})$ is the $6\times6$ matrix of correlations. Since the only non-zero correlations are of the type $\langle d_x(t) d_x(t)\rangle$ or  $\langle d_x(t)d_x(t+\tau)$, this matrix is band-diagonal, with ${1\over3}\rho_{mn}(0)$ on the main diagonal and ${1\over 3}\rho_{mn}(\tau)$ on the upper and lower diagonals two removed from the main diagonal. The higher joint probabilities have similar forms, but with enlarged vector ${\bf \Delta}$ and enlarged matrix ${\bf C}$ \cite{fox:1978}.
From the Gaussian nature of the process, and by Wick's theorem, the four-point correlation function  $\rho^{(4)}_{mn}(\tau, \tau^{\prime}, \tau^{\prime\prime}) = \langle {\bf d}_{mn}(t)\cdot{\bf d}_{mn}(t + \tau){\bf d}_{mn}(t + \tau^{\prime})\cdot{\bf d}_{mn}(t + \tau^{\prime\prime})\rangle$ may be calculated as sums of  products of the two point correlation function $\rho_{mn}(\tau)$. It is important to emphasis that Xie {\it et al}  have explicitly checked that the higher order correlations obey Wick's theorem and haver thereby confirmed that the distance fluctuations are Gaussian. 

We have been able to characterise the stochastic process governing the distance fluctuations in terms of its correlation function and the joint probabilities. It is of interest to ask if there is an effective description, in terms of a suitable Langevin equation, for the distance fluctuation ${\bf d}_{mn}(t)$ itself.  In other words, is it possible to construct a Langevin equation for ${\bf d}_{mn}(t)$, given its correlation function  $\rho_{mn}(\tau)$ ? Fox \cite{fox:1978} has provided a solution to this inverse problem in terms of a {\it linear} Langevin equation with a memory kernel and Gaussian, coloured noise. The effective Langevin equation for the superposed OUP, then,  is
\begin{equation}
{d\over dt} {\bf d}_{mn}(t) = -\int_0^t D_{mn}(t-t^{\prime}){\bf d}_{mn}(t^{\prime}) + {\bf f}_{mn}(t)
\end{equation}
where the noise is Gaussian with mean $\langle {\bf f}_{mn}(t)\rangle = 0$ and variance $\langle f_{mni}(t)f_{mnj}(t+\tau)\rangle = 2k_BTD_{mn}(\tau)$. It can be shown \cite{fox:1978} that this Langevin equation yields the correlation function $\rho_{mn}(\tau)$ and the full heirarchy of joint probabilities described above, provided the Laplace transform of the diffusion kernel $D_{mn}(s) = \int_0^{\infty}d\tau \exp(-s\tau)D_{mn}(\tau)$ satisfies
\begin{equation}
\rho_{mn}(s) = {\rho_{mn}(0)\over s + D_{mn}(s)},
\end{equation}
where $\rho_{mn}(s)$ is the Laplace transform of $\rho_{mn}(\tau)$. An explicit expression for $D_{mn}(s)$ can be obtained by Laplacing transforming Eq.\ref{eq:corr}, replacing the summation by an integration, and inserting the expression for $\rho_{mn}(s)$ in the equation above. For $m=0, n=N$, we get for $D(s) = D_{0N}(s)$,
\begin{equation}
D(s) = \sqrt{s \over \tau_1}{{\pi\over 2} - \tan^{-1}({1\over\sqrt{s\tau_1}}) \over 1 - {1\over\sqrt{s\tau_1}}[{\pi\over 2} -\tan^{-1}({1\over\sqrt{s\tau_1}})]}
\end{equation}
From this, we may obtain an effective friction kernel $\zeta(s)$ using the generalised Stokes-Einstein relation $D(s) = k_BT/\zeta(s)$. This effective friction kernel recieves contributions from individual dissipative effects of all the Rouse modes and is the source of memory. For $s\tau_1 \ll 1$, which corresponds to times much greater than the longest relaxation time, $\zeta(s) \rightarrow \tau_1$, which implies that the memory kernel is proportional to $\delta(t)$. Reassuringly, the Markovian limit is reproduced for times much longer than the longest time scale in the problem. On the other hand, for $s\tau_1 \gg 1$, $\zeta(s)\rightarrow {2\over\pi}\sqrt{\tau_1/s}$, implying that $\zeta(t)$ is proportional to $(\tau_1/t)^{1/2}$. Thus, for times smaller than the longest relaxation time, the memory kernel shows a power law decay in time with an exponent of $-{1\over 2}$. Correspondingly, the normalized correlation function $C(s) = \rho(s)/\rho(0)$ is 
\begin{equation}
C(s) =  \frac{1}{s}\left[ {1 - {1\over\sqrt{s\tau_1}}[{\pi\over 2} -\tan^{-1}({1\over\sqrt{s\tau_1}})]}\right],
\end{equation}
where $\rho(s) =\rho_{0N}(s)$. For $s\tau_1 \gg 1$, $C(s) \rightarrow 1/s$, which implies that at short times $C(t) = 1$. 
In the opposite limit of $s\tau_1 \ll 1$, $C(s) \approx \tau_1/(1+s\tau_1)$, which implies that at long times $C(t)$ is proportional to $\exp(-t/\tau_1)$ and decays as a single exponential, again suggesting that the fluctuations become Markovian at long times.  These asymptotes can be understood by a simple scaling argument. The normalized correlation function is given by $C(t) = \sum_{p,odd} e^{-p^2 t/\tau_1}/p^2$. It is evident from the expression for $C(t)$ that for  times $t\gtrsim \tau_1$, when all but the longest modes have relaxed, $C(t)$ decays as a single exponential. For times $t \gg \tau_1/p^2$, all modes above the $p$-th mode have already relaxed and do not contribute to the summation. This implies that at any time $t$, modes $p = 1, 2 \ldots p^{\star}(t)$ contribute to the summation, where $p^{\star}(t) \approx (\tau_1/t)^{1/2}$. The slow decay of this maximum mode number shows up as a non-Markovian effect and is ultimately responsible for the power-law dependence seen in the frictional memory kernel.

Our work has precursors in the contribution of Kou and Xie \cite{kou:2004} who proposed a one-dimensional phenomenological Langevin equation incorporating fractional Gaussian noise. However, no microscopic basis was provided for memory kernel or for the form of the noise. Starting from a microscopic description, we obtain a vector Langevin equation with memory, obtain an explicit expression for the memory kernel, and show that this has a power law decay. The noise in our Langevin equation is not   a phenomenological fractional Gaussian noise, but is fully specified from the microscopics. Debnath {\it et al} \cite{debnath:2005} use a semi-flexible model of polymer dynamics to explain the experimental data. Our work shows that semiflexibility is not necessary to understand those aspects of protein dynamics probed by the fluorescence experiments. A similar approach is that of  Tang and Marcus \cite{tang:2006} for a flexible polymer, but the experimental results can be obtained only by assuming a certain disorder along the chain and averaging over the disorder. Our work shows that the heterogeneity of the protein residues is not necessary to explain the experimental results. None of the above have elucidated the appearence of non-Markovian behaviour from a simple superposition of Markovian fluctuations as is done here.

In summary, we have presented a mechanism where a superposition of the Markovian dynamics of the normal modes of the protein conformation gives rise to non-Markovian fluctuations of the distance between two points in the protein. The model provides an accurate fit to experimental data for both two-time and four-time correlation functions, exhibits a symmetry of these correlations functions found experimentally, reproduces the power-law decay of the frictional memory kernel, and clarifies how non-Markovian behaviour arises in protein dynamics. Given the non-specific nature of our model, we believe that  non-Markovian fluctuations should be seen universally in all biopolymers and not only in proteins. Our model can be extended to include more detailed descriptions of the protein normal modes and their relaxation mechanism. Using our model, several quantities of interest in fluorescence microscopy like the survival and first passage times of the distance may be calculated. Work is underway to explore these possibilities. 

This work was first presented at the 2008 Biophysics Summer School, at the Harishchandra Research Institute, Allahabad. RA wishes to thank the organisers and participants for useful comments and suggestions.  

\bibliography{proteinfluct}

\begin{thebibliography}{11}
\expandafter\ifx\csname natexlab\endcsname\relax\def\natexlab#1{#1}\fi
\expandafter\ifx\csname bibnamefont\endcsname\relax
  \def\bibnamefont#1{#1}\fi
\expandafter\ifx\csname bibfnamefont\endcsname\relax
  \def\bibfnamefont#1{#1}\fi
\expandafter\ifx\csname citenamefont\endcsname\relax
  \def\citenamefont#1{#1}\fi
\expandafter\ifx\csname url\endcsname\relax
  \def\url#1{\texttt{#1}}\fi
\expandafter\ifx\csname urlprefix\endcsname\relax\def\urlprefix{URL }\fi
\providecommand{\bibinfo}[2]{#2}
\providecommand{\eprint}[2][]{\url{#2}}

\bibitem[{\citenamefont{Lu et~al.}(1998)\citenamefont{Lu, Xun, and
  Xie}}]{lu:1998}
\bibinfo{author}{\bibfnamefont{H.~P.} \bibnamefont{Lu}},
  \bibinfo{author}{\bibfnamefont{L.-Y.} \bibnamefont{Xun}}, \bibnamefont{and}
  \bibinfo{author}{\bibfnamefont{X.~S.} \bibnamefont{Xie}},
  \bibinfo{journal}{Science} \textbf{\bibinfo{volume}{282}},
  \bibinfo{pages}{1877} (\bibinfo{year}{1998}).

\bibitem[{\citenamefont{van Oijen et~al.}(2003)\citenamefont{van Oijen,
  Blainey, Crampton, Richardson, Ellenberger, and Xie}}]{oijen:2003}
\bibinfo{author}{\bibfnamefont{A.~M.} \bibnamefont{van Oijen}},
  \bibinfo{author}{\bibfnamefont{P.}~\bibnamefont{Blainey}},
  \bibinfo{author}{\bibfnamefont{D.~J.} \bibnamefont{Crampton}},
  \bibinfo{author}{\bibfnamefont{C.~C.} \bibnamefont{Richardson}},
  \bibinfo{author}{\bibfnamefont{T.}~\bibnamefont{Ellenberger}},
  \bibnamefont{and} \bibinfo{author}{\bibfnamefont{X.~S.} \bibnamefont{Xie}},
  \bibinfo{journal}{Science} \textbf{\bibinfo{volume}{301}},
  \bibinfo{pages}{1235} (\bibinfo{year}{2003}).

\bibitem[{\citenamefont{Yang et~al.}(2003)\citenamefont{Yang, Luo,
  Karnchanaphanurach, Louie, Rech, Cova, Xun, and Xie}}]{yang:2003}
\bibinfo{author}{\bibfnamefont{H.}~\bibnamefont{Yang}},
  \bibinfo{author}{\bibfnamefont{G.}~\bibnamefont{Luo}},
  \bibinfo{author}{\bibfnamefont{P.}~\bibnamefont{Karnchanaphanurach}},
  \bibinfo{author}{\bibfnamefont{T.-M.} \bibnamefont{Louie}},
  \bibinfo{author}{\bibfnamefont{I.}~\bibnamefont{Rech}},
  \bibinfo{author}{\bibfnamefont{S.}~\bibnamefont{Cova}},
  \bibinfo{author}{\bibfnamefont{L.}~\bibnamefont{Xun}}, \bibnamefont{and}
  \bibinfo{author}{\bibfnamefont{X.~S.} \bibnamefont{Xie}},
  \bibinfo{journal}{Science} \textbf{\bibinfo{volume}{302}},
  \bibinfo{pages}{262} (\bibinfo{year}{2003}).

\bibitem[{\citenamefont{Kou and Xie}(2005)}]{kou:2004}
\bibinfo{author}{\bibfnamefont{S.~C.} \bibnamefont{Kou}} \bibnamefont{and}
  \bibinfo{author}{\bibfnamefont{X.~S.} \bibnamefont{Xie}},
  \bibinfo{journal}{Phys. Rev. Lett} \textbf{\bibinfo{volume}{93}},
  \bibinfo{pages}{180603} (\bibinfo{year}{2005}).

\bibitem[{\citenamefont{Min et~al.}(2005)\citenamefont{Min, Luo, Cherayil, Kou,
  and Xie}}]{min:2005}
\bibinfo{author}{\bibfnamefont{W.}~\bibnamefont{Min}},
  \bibinfo{author}{\bibfnamefont{G.}~\bibnamefont{Luo}},
  \bibinfo{author}{\bibfnamefont{B.~J.} \bibnamefont{Cherayil}},
  \bibinfo{author}{\bibfnamefont{S.~C.} \bibnamefont{Kou}}, \bibnamefont{and}
  \bibinfo{author}{\bibfnamefont{X.~S.} \bibnamefont{Xie}},
  \bibinfo{journal}{Phys. Rev. Lett} \textbf{\bibinfo{volume}{94}},
  \bibinfo{pages}{198302} (\bibinfo{year}{2005}).

\bibitem[{\citenamefont{Kampen}(1981)}]{vankampen:1981}
\bibinfo{author}{\bibfnamefont{N.~G.~V.} \bibnamefont{Kampen}},
  \emph{\bibinfo{title}{Stochastic Processes in Physics and Chemistry}}
  (\bibinfo{publisher}{North Holland}, \bibinfo{address}{Amsterdam, The
  Netherlands}, \bibinfo{year}{1981}).

\bibitem[{\citenamefont{Banavar et~al.}(2005)\citenamefont{Banavar, Hoang, and
  Maritan}}]{banavar:2005}
\bibinfo{author}{\bibfnamefont{J.~R.} \bibnamefont{Banavar}},
  \bibinfo{author}{\bibfnamefont{T.~X.} \bibnamefont{Hoang}}, \bibnamefont{and}
  \bibinfo{author}{\bibfnamefont{A.}~\bibnamefont{Maritan}},
  \bibinfo{journal}{The Journal of Chemical Physics}
  \textbf{\bibinfo{volume}{122}}, \bibinfo{pages}{234910}
  (\bibinfo{year}{2005}).

\bibitem[{\citenamefont{Doi and Edwards}(1986)}]{doi:1986}
\bibinfo{author}{\bibfnamefont{M.}~\bibnamefont{Doi}} \bibnamefont{and}
  \bibinfo{author}{\bibfnamefont{S.}~\bibnamefont{Edwards}},
  \emph{\bibinfo{title}{The theory of polymer dynamics}}
  (\bibinfo{publisher}{Clarendon Press, Oxford}, \bibinfo{year}{1986}).

\bibitem[{\citenamefont{Fox}(1978)}]{fox:1978}
\bibinfo{author}{\bibfnamefont{R.~F.} \bibnamefont{Fox}},
  \bibinfo{journal}{Physics Reports} \textbf{\bibinfo{volume}{48}},
  \bibinfo{pages}{179} (\bibinfo{year}{1978}).

\bibitem[{\citenamefont{Debnath et~al.}(2005)\citenamefont{Debnath, Min, Xie,
  and Cherayil}}]{debnath:2005}
\bibinfo{author}{\bibfnamefont{P.}~\bibnamefont{Debnath}},
  \bibinfo{author}{\bibfnamefont{W.}~\bibnamefont{Min}},
  \bibinfo{author}{\bibfnamefont{X.~S.} \bibnamefont{Xie}}, \bibnamefont{and}
  \bibinfo{author}{\bibfnamefont{B.~J.} \bibnamefont{Cherayil}},
  \bibinfo{journal}{The Journal of Chemical Physics}
  \textbf{\bibinfo{volume}{123}}, \bibinfo{pages}{204903}
  (\bibinfo{year}{2005}).

\bibitem[{\citenamefont{Tang and Marcus}(2006)}]{tang:2006}
\bibinfo{author}{\bibfnamefont{J.}~\bibnamefont{Tang}} \bibnamefont{and}
  \bibinfo{author}{\bibfnamefont{R.~A.} \bibnamefont{Marcus}},
  \bibinfo{journal}{Phys. Rev. E} \textbf{\bibinfo{volume}{73}},
  \bibinfo{pages}{022102} (\bibinfo{year}{2006}).

\end{thebibliography}

\end{document}